\documentclass[showpacs,amsmath,nofootinbib,amssymb,epsfig]{revtex4}

\usepackage{amsfonts,amssymb,amsmath}
\usepackage[dvips]{epsfig}
\usepackage[T1]{fontenc}
\usepackage[latin1]{inputenc}
\usepackage{graphicx}
\usepackage[english]{babel}
\usepackage{amsmath}
\usepackage{amssymb}
\usepackage{amsfonts}
\usepackage{color}

\begin{document}

\title{\bf Einstein Static Universe in Rastall Theory of Gravity}
\author{F. Darabi $^{1}$}
\email{f.darabi@azaruniv.ac.ir}\affiliation{$^{1}$Department of Physics, Azarbaijan Shahid Madani University , Tabriz, 53714-161 Iran\\$^{2}$Research Institute for Astronomy and Astrophysics of Maragha (RIAAM),
Maragha 55134-441, Iran}
\author{K. Atazadeh$^{1,2}$}
\email{atazadeh@azaruniv.ac.ir}\affiliation{$^{1}$Department of Physics, Azarbaijan Shahid Madani University , Tabriz, 53714-161 Iran\\$^{2}$Research Institute for Astronomy and Astrophysics of Maragha (RIAAM),
Maragha 55134-441, Iran}
\author{Y. Heydarzade$^{1}$}
\email{heydarzade@azaruniv.ac.ir}\affiliation{$^{1}$Department of Physics, Azarbaijan Shahid Madani University , Tabriz, 53714-161 Iran\\$^{2}$Research Institute for Astronomy and Astrophysics of Maragha (RIAAM),
Maragha 55134-441, Iran}

\date{\today}

\begin{abstract}
We investigate stability of the Einstein static solution against homogeneous scalar, vector and tensor perturbations in the context of Rastall theory of gravity. We show that this solution in the presence of perfect fluid and vacuum energy originating from conformally-invariant fields is stable. Using
the fix point method and  taking linear homogeneous perturbations, we find that the scale factor of Einstein static universe for closed deformed isotropic and homogeneous FLRW universe depends on the coupling parameter  $\lambda$ between the energy-momentum tensor and the gradient of Ricci scalar. Thus, in the present model and in presence of vacuum energy, our universe can stay at the Einstein static state past-eternally, which means that the big bang singularity may be resolved successfully in the context of Einstein static universe in Rastall theory.
\end{abstract}
\maketitle

\section{Introduction}
Covariant conservation of  energy-momentum tensor is one of the basic elements of Einstein's general theory of relativity (GR) which leads, via  Noether symmetry theorem,  to the conservation of some globally defined physical quantities. These conserved quantities appear as the integrals of the components
of  energy-momentum tensor over appropriate space-like hypersurfaces. These space-like hypersurfaces admit at
least one of the Killing vectors of the background spacetime as their normal
vectors. In this way, the total rest
energy/mass of a physical system is conserved in the context of GR. On the other hand, some GR based new
modified theories of gravity have been proposed by relaxing the condition of covariant energy-momentum conservation.
One of these  modified theories  of gravity was proposed by P. Rastall in 1972
\cite{1, 2}. The remarkable point in this theory is that the usual conservation law expressed by the null divergence of the energy-momentum
tensor, i.e $T^{\mu\nu}_{~~~;\mu} = 0$, is questioned. Instead, a non-minimal coupling of matter fields to geometry is considered
where the divergence of $T_{\mu\nu}$ is proportional to the gradient of the Ricci scalar, i.e $T^{\mu\nu}_{~~~;\mu} =$ $\lambda R^{,\nu}$, such
that the usual conservation law is recovered in the flat spacetime. This can be interpreted as a direct
consequence of  Mach principle indicating that the inertia of a local mass   depends on
the global mass and energy distribution in the universe \cite{3}. The main argument in favor of such a proposal
is that the usual conservation law on $T^{\mu\nu}$ is tested only in the flat Minkowski space-time or at most in
a gravitational weak field limit. Indeed, Rastall theory reproduces a phenomenological way for distinguishing
characteristic features of quantum effects in gravitational systems, i.e the violation of  classical conservation laws
\cite{4, 5, 6}, which is also reported in $f(R, T )$ \cite{7} and $f(R,{\cal L}_{m})$ \cite{8} theories, where $R, T$ and ${\cal L}_{m}$ are the
Ricci scalar, trace of the energy-momentum tensor and the Lagrangian of the matter sector, respectively.
Also, the violation of covariant conservation of  energy-momentum tensor  $T^{\mu\nu}_{~~~;\mu} \neq 0$ is phenomenologically confirmed by the particle creation process in cosmology
\cite{9, 10, 11, 12, 13, 14, 15, 16}.  In this regard, the Rastall theory can be considered as a good candidate for classical formulation of the particle creation through its non-minimal coupling \cite{12, 17}. Moreover, some
astrophysical analysis including the evolution of  neutron stars and cosmological data do not rule out the Rastall
theory \cite{18, 19, 20}. Specially, in \cite{18} it is shown that the restrictions on the Rastall geometric
parameters are of  order  $\leq$ 1\%\ with respect to the corresponding value of  Einstein GR. In other
words, the results in \cite{18} confirm that the Rastall theory is a viable theory in the sense that the deviation
of any extended theory of gravity from  standard GR must be so weak that
it can pass the solar system tests.
Some other studies on the various aspects of this theory in the context of current accelerated expansion phase
of the universe, as well as other cosmological problems, can be found in \cite{12, 21, 22, 23, 24, 25, 26, 27, 28}.

Besides the potential aspects of Rastall theory in the cosmological context
to describe the current acceleration of the universe,
a new cosmological scenario
in the framework of Einstein's general relativity,  so  called  emergent universe, was introduced in  \cite{33,44} to remove the initial singularity. According to this scenario, the universe might have been originated from an Einstein static state rather
than a big bang singularity. This scenario is completely different from
the  ``static universe'' originally introduced by Einstein to describe the
large scale universe at that time. Rather, this scenario describes the very early
universe by supporting a past-eternal inflationary model in which the Big-Bang singularity is removed and the horizon
problem is solved before the beginning of inflation.  The inflationary universe in this scenario emerges from a
very small size static universe containing the seeds for the development of
the microscopic universe. However, because of the existence of varieties of perturbations,
such as the quantum fluctuations, this cosmological model is unstable and
suffers
from a fine-tuning problem which can be amended by
modifications of the cosmological equations of general relativity.
In order to overcome the un-stability problem,  the scenario of Einstein static universe has been explored
in the context of different modified theories of gravity  \cite{Carneiro, Mulryne, Parisi, Wu,
Lidsey, Bohmer2007, Seahra, Bohmer2009, Barrow,
Barrow2009, Clifton,Boehmer2010, Boehmer20093, Wu20092,
Odrzywolek}, from loop quantum gravity \cite{Mulryne, Parisi, Wu} to $f(R)$ gravity  \cite{Boehmer2010} and $f(T)$ gravity \cite{ft}, from Horava-Lifshitz gravity \cite{Wu20092,Odrzywolek} to brane gravity \cite{Lidsey} and massive gravity \cite{mass}.

In this paper, motivated by the above mentioned research for finding a viable modified theory
of gravity which i) can potentially describe the accelerated expansion of
the universe and ii)  can describe an emergent universe within the scenario of Einstein static universe to resolve the initial singularity problem, we
study  the ``Rastall theory'' in the
framework of ``Einstein static universe''. We
consider the stability of  Einstein static universe in the Friedmann-Lema\^{\i}tre-Robertson-Walker (FRLW) space-time  in sections II and III only with matter.  In Section V, we present an analysis of the
equilibrium of Einstein solution of the model in the presence of matter and the vacuum energy. Next, in the section VI, from the perspective of fix point method, we consider a numerical example, in which the energy content
contains the relativistic matter plus a vacuum term with negative pressure.
The paper ends with a brief conclusions in Section VII.

\section{Rastall Theory and Its Friedmann Equations}
Rastall challenged the conservation of the energy-momentum tensor $T^{\mu\nu}_{~~~;\mu} = 0$ in curved spacetime, and proposed a new theory for gravity by assuming
$T^{\mu\nu}_{~~~;\mu} =$ $\lambda R^{,\nu}$, where $\lambda$ is a constant which should be determined from observations and other parts of physics \cite{1}. Thus, for a spacetime metric $g_{\mu\nu}$, the corresponding gravitational field equations can be written as
\begin{eqnarray}
G_{\mu\nu}+ k\lambda g_{\mu\nu}R = kT_{\mu\nu},
\end{eqnarray}
where $G_{\mu\nu}$ and $T_{\mu\nu}$ are the Einstein and energy-momentum tensors, respectively \cite{1}. Also, $R$ is Ricci scalar,
and $k$ is also gravitational constant in Rastall theory. It can be seen that for $\lambda = 0$ and $k = 8\pi G$ the Einstein field equations are recovered wherever
$T^{\mu\nu}_{~~~;\mu} = 0$. By taking a cosmological background described by FLRW metric
\begin{eqnarray}
ds^2=-dt^2+a(t)^2\left[\frac{dr^2}{1-\kappa r^2}+r^2d\Omega^2\right],
\end{eqnarray}
where $a(t)$ and $\kappa$ denote the scale factor and the curvature parameter, respectively, while $\kappa = -1, 0, 1$ denotes the open,
flat and closed universes, respectively.
In this case, using the Rastall field equation, we find the Friedmann equations for a perfect fluid as \cite{yuan}
\begin{eqnarray}\label{fr}
3\left(1-4k\lambda\right)H^2-6k\lambda\dot H+3\left( 1-2k\lambda  \right)\frac{\kappa}{a^2}&=&
k\rho,\nonumber\\
3\left(1-4k\lambda\right)H^2+2(1-3k\lambda)\dot H+\left( 1-6k\lambda  \right)\frac{\kappa}{a^2}&=&-k p.
\end{eqnarray}
Also, the conservation equations for the matter component is
given by
\begin{eqnarray}\label{ma}
\frac{1-3k\lambda}{1-4k\lambda}\dot{\rho}-\frac{3k\lambda}{1-4k\lambda}\dot{p}+ 3H(\rho + p) =0\,.
\end{eqnarray}
\section{ Einstein Static Universe and Stability
Analysis}
\subsection{ESU Versus the Scalar Perturbations}
 For an ESU, the condition $\ddot{a}=0=\dot{a}$ is required. Then, regarding the Friedmann
 equations (\ref{fr}), we have the following equations for the ESU
 \begin{eqnarray}\label{esu}
3\left( 1-2k\lambda  \right)\frac{\kappa}{a_0^2}&=&k\rho_0  ,\nonumber\\
\left( 1-6k\lambda  \right)\frac{\kappa}{a_0^2}&=&-kP_{0}.
\end{eqnarray}
From the above equations we obtain
\begin{eqnarray}\label{ome}
\omega=\frac{\left( -1+6k\lambda  \right)}{3\left( 1-2k\lambda  \right)}.
\end{eqnarray}
In what follows, we consider  the linear homogeneous scalar perturbations of
equations in (\ref{fr}) around
the ESU described by the equations in (\ref{esu}) and
explore their stability against these perturbations. The  perturbation in the cosmic scale factor $a(t)$ and the energy density
$\rho(t)$ can be considered as
\begin{eqnarray}\label{bbmm}
&&a(t)\rightarrow a_{0}(1+\delta a(t)),\nonumber\\
&&\rho(t)\rightarrow \rho_{0}(1+\delta \rho(t)).
\end{eqnarray}
Substituting these perturbations in the first equation in the set of equations (\ref{fr})
and linearizing the result
gives the following equation

\begin{equation}\label{1}
-6k\lambda \ddot\delta a +3(1-2k\lambda)\frac{\kappa}{a_0^2}\left(1-2\delta a\right)=k\rho_0(1+\delta \rho),
\end{equation}
Then, using the equations (\ref{esu}) describing an ESU, we have
 \begin{equation}\label{bbn}
-6k\lambda \ddot\delta a -6(1-2k\lambda)\frac{\kappa}{a_0^2}\delta a=k\rho_0\delta \rho.
\end{equation}
Similarly, for the   second equation in the set of equations (\ref{fr})
and linearizing the result, we arrive at
 \begin{equation}\label{2}
2(1-3k\lambda) \ddot\delta a -2(1-6k\lambda)\frac{\kappa}{a_0^2}\delta a=-k\omega\rho_0\delta \rho.
\end{equation}
Substituting $k\rho_0\delta \rho$ from (\ref{1}) in (\ref{2}) leads to the
following equation
 \begin{equation}\label{3}
 \ddot\delta a -\left(\frac{2-12k\lambda(1+\omega)+6\omega}{2-6k\lambda(1+\omega)}\right)\frac{\kappa}{a_0^2}\delta a=0.
\end{equation}
Substituting $\omega$ from (\ref{ome}) into this equation, we arrive at
 \begin{equation}\label{4}
\ddot \delta a =0.
\end{equation}
Then, we have no oscillator equation and consequently, ESU is unstable versus the scalar perturbations in the context of Rastall theory.
\subsection{ESU Versus the Vector and Tensor Perturbations}
In the cosmological setup, the vector perturbations of a perfect fluid with energy density $\rho$
and barotropic equation of state  are governed by the co-moving dimensionless {\it vorticity} defined as ${\varpi}_a=a{\varpi}$ \cite{pert}.
The vorticity modes satisfy the following propagation equation \cite{pert}
\begin{equation}\label{v}
\dot{\varpi}_{\kappa}+(1-3c_s^2)H{\varpi}_{\kappa}=0,
\end{equation}
where $c_s^2=dp/d\rho$ and $H$  are the sound speed and the Hubble parameter,
respectively.
This equation is valid in our treatment of ESU in the
framework of the Rastall theory through the
modified   Friedmann equations given in  (\ref{fr}) . For the ESU with $H=0$, the propagation equation (\ref{v}) reduces to
\begin{equation}\label{}
\dot{\varpi}_{\kappa}=0.
\end{equation}
This  indicates  that the initial vector perturbations remain frozen. Thus,
 we have a neutral stability versus the vector perturbations.

Tensor perturbations, namely gravitational-wave perturbations, of a perfect
fluid is described by the co-moving dimensionless transverse-traceless shear $\Sigma_{ab}=a\sigma_{ab}$, whose modes satisfy the following propagation
equation
\begin{equation}\label{438}
\ddot\Sigma_{\kappa}+3H\dot\Sigma_{\kappa}+\left[\frac{\mathcal{K}^2}{a^2}
+\frac{2\kappa}{a^2}-\frac{8\pi}{3}(1+3\omega)\rho\right]\Sigma_{\kappa}=0,
\end{equation}
where $\mathcal{K}$ is the co-moving index ($D^2\rightarrow -\mathcal{K}^2/a^2$ in which $D^2$ is
the covariant spatial Laplacian)\cite{pert}.
For the ESU,
using the equations (\ref{esu}), (\ref{bbn}) and (\ref{2}), this equation  reduces to the following form
\begin{equation}\label{439}
\ddot\Sigma_{\kappa}+\frac{1}{ a_0^2}\left[\mathcal{K}^{2}+2\kappa(1-16\pi\lambda)\right]\Sigma_{\kappa}=0.
\end{equation}
Then, in order to have stable modes against the tensor perturbations, the following condition
is required
\begin{equation}
\mathcal{K}^{2}+2\kappa(1-16\pi\lambda)>0.
\end{equation}
This inequality gives a constraint on the geometric parameter $\lambda$ of the Rastall
theory to have a stable ESU against the tensor perturbations as
\begin{equation}
\lambda<\frac{1}{16\pi}\left( 1+\frac{\mathcal{K}^{2}}{2\kappa} \right).
\end{equation}

 \section{ Rastall Field Equations with the Vacuum Energy Term}

In this case, using the Rastall field equation, we find the Friedmann equations for a perfect fluid as \cite{yuan}
\begin{eqnarray}\label{fr}
3\left(1-4k\lambda\right)H^2-6k\lambda\dot H+3\left( 1-2k\lambda  \right)\frac{\kappa}{a^2}&=&k\left(\rho+\rho_\Lambda\right)=
k\left(\rho+\frac{A}{a^4}\right),\nonumber\\
3\left(1-4k\lambda\right)H^2+2(1-3k\lambda)\dot H+\left( 1-6k\lambda  \right)\frac{\kappa}{a^2}&=&-k\left(p+p_\Lambda\right)
=-k\left(\omega\rho+\omega_\Lambda \frac{A}{a^4} \right) .
\end{eqnarray}
Also, the conservation equation in the presence of vacuum energy is
given by
\begin{eqnarray}\label{ma}
\frac{1-3k\lambda}{1-4k\lambda}(\dot{\rho}+\dot{\rho}_{\Lambda})-\frac{3k\lambda}{1-4k\lambda}(\dot{p}+\dot{p}_{\Lambda})+ 3H(\rho+ \rho_{\Lambda}+ p+p_{\Lambda}) =0\,.
\end{eqnarray}
\section{ Einstein Static Universe and Stability
Analysis}
\subsection{ESU Versus the Scalar Perturbations}
 For an ESU, the condition $\ddot{a}=0=\dot{a}$ is required. Then, regarding the Friedmann
 equations (\ref{fr}), we have the following equations for the ESU
 \begin{eqnarray}\label{esu}
3\left( 1-2k\lambda  \right)\frac{\kappa}{a_0^2}&=&k\left(\rho_0 +\frac{A}{a_0^4}\right) ,\nonumber\\
\left( 1-6k\lambda  \right)\frac{\kappa}{a_0^2}&=&-k\left(p_{0}+p_{0\Lambda}\right)=-k\left(\omega\rho_{0}+\omega_\Lambda\rho_{0\Lambda}\right).
\end{eqnarray}

In what follows, we consider  the linear homogeneous scalar perturbations of
equations in (\ref{fr}) around
the ESU described by the equations in (\ref{esu}) and
explore their stability against these perturbations. The  perturbation in the cosmic scale factor $a(t)$ and the energy density
$\rho(t)$ can be considered as
\begin{eqnarray}\label{bbmm}
&&a(t)\rightarrow a_{0}(1+\delta a(t)),\nonumber\\
&&\rho(t)\rightarrow \rho_{0}(1+\delta \rho(t)).
\end{eqnarray}
Substituting these perturbations in the first equation in the set of equations (\ref{fr})
and linearizing the result
gives the following equation

\begin{equation}\label{1}
-6k\lambda \ddot\delta a +3(1-2k\lambda)\frac{\kappa}{a_0^2}\left(1-2\delta a\right)=k\left[\rho_0(1+\delta \rho)+\frac{A}{a_0^4}(1-4\delta a)\right],
\end{equation}
Then, using the equations (\ref{esu}) describing an ESU, we have
 \begin{equation}\label{bbn}
-6k\lambda \ddot\delta a -6(1-2k\lambda)\frac{\kappa}{a_0^2}\delta a=k\left[\rho_0\delta \rho-\frac{4A}{a_0^4}\delta a\right].
\end{equation}
Similarly, for the   second equation in the set of equations (\ref{fr})
and linearizing the result, we arrive at
 \begin{equation}\label{2}
2(1-3k\lambda) \ddot\delta a -2(1-6k\lambda)\frac{\kappa}{a_0^2}\delta a=-k\left[\omega\rho_0\delta \rho -\frac{4\omega_\Lambda A}{a_0^4}\delta a\right].
\end{equation}
Substituting $k\rho_0\delta \rho$ from (\ref{bbn}) in (\ref{2}) leads to the
following equation
 \begin{equation}\label{3}
\left(2-6k\lambda(1+\omega)\right) \ddot\delta a +\left[\left(-2(1+3\omega)+12k\lambda(1+\omega)\right)\frac{\kappa}{a_0^2}+\frac{4kA}{a_0^4}(\omega-\omega_\Lambda)\right]\delta a=0.
\end{equation}
Using both equations (\ref{esu}), one may rewrite this equation in the following
form \begin{equation}\label{4}
\ddot \delta a +\frac{\kappa}{a_0^2}\left[ \frac{1+3\omega-6k\lambda(1+\omega)}{1-3k\lambda(1+\omega)}  \right]\delta a=0.
\end{equation}
Then, we have the  oscillator equations for the closed and open universes governed by the following conditions
 \begin{equation}\label{cons}
\begin{cases}\frac{1+3\omega-6k\lambda(1+\omega)}{1-3k\lambda(1+\omega)}>0~~ &\kappa>0 ,
\\
\\
\frac{1+3\omega-6k\lambda(1+\omega)}{1-3k\lambda(1+\omega)}<0~~ &\kappa>0,
\end{cases} \end{equation}
while for the flat universe, i.e $\kappa=0$, there is no stable ESU.
\subsection{ESU Versus the Vector and Tensor Perturbations}
In the cosmological setup, the vector perturbations of a perfect fluid with energy density $\rho$
and barotropic equation of state  are governed by the co-moving dimensionless {\it vorticity} defined as ${\varpi}_a=a{\varpi}$ \cite{pert}.
The vorticity modes satisfy the following propagation equation \cite{pert}
\begin{equation}\label{v}
\dot{\varpi}_{\kappa}+(1-3c_s^2)H{\varpi}_{\kappa}=0,
\end{equation}
where $c_s^2=dp/d\rho$ and $H$  are the sound speed and the Hubble parameter,
respectively.
This equation is valid in our treatment of ESU in the
framework of the Rastall theory through the
modified   Friedmann equations given in  (\ref{fr}) . For the ESU with $H=0$, the propagation equation (\ref{v}) reduces to
\begin{equation}\label{}
\dot{\varpi}_{\kappa}=0.
\end{equation}
This  indicates  that the initial vector perturbations remain frozen. Thus,
 we have a neutral stability versus the vector perturbations.

Tensor perturbations, namely gravitational-wave perturbations, of a perfect
fluid is described by the co-moving dimensionless transverse-traceless shear $\Sigma_{ab}=a\sigma_{ab}$, whose modes satisfy the following propagation
equation
\begin{equation}\label{438}
\ddot\Sigma_{\kappa}+3H\dot\Sigma_{\kappa}+\left[\frac{\mathcal{K}^2}{a^2}
+\frac{2\kappa}{a^2}-\frac{8\pi}{3}(1+3\omega)\rho\right]\Sigma_{\kappa}=0,
\end{equation}
where $\mathcal{K}$ is the co-moving index ($D^2\rightarrow -\mathcal{K}^2/a^2$ in which $D^2$ is
the covariant spatial Laplacian)\cite{pert}.
For the ESU in Rastall theory,
using the equations (\ref{esu}), (\ref{bbn}) and (\ref{2}), this equation  reduces to the following form
\begin{equation}\label{439}
\ddot\Sigma_{\kappa}+\frac{1}{ a_0^2}\left[\mathcal{K}^{2}+2\kappa(1-16\pi\lambda)+\frac{A}{a_0^2}(1+3\omega_\Lambda)\right]\Sigma_{\kappa}=0.
\end{equation}
Then, in order to have stable modes against the tensor perturbations, the following condition
is required
\begin{equation}
\mathcal{K}^{2}+2\kappa(1-16\pi\lambda)+\frac{A}{a_0^2}(1+3\omega_\Lambda)>0.
\end{equation}
This inequality gives a constraint on the geometric parameter $\lambda$ of the Rastall
theory to have a stable ESU against the tensor perturbations as
\begin{equation}
\lambda<\frac{1}{16\pi}\left( 1+\frac{1}{2\kappa}\left( \mathcal{K}^{2}+\frac{A}{a_0^2}(1+3\omega_\Lambda) \right) \right).
\end{equation}
One may rewrite this condition using the equations (\ref{esu}) in the following
form
\begin{equation}
\rho_0a_0^2\left(1+3\omega\right)<\mathcal{K}^{2}+2\kappa\left(1-6\lambda(1-\pi)\right),
\end{equation}
which represents that how the geometric parameter of the Rastall theory affects the matter content and size of the initial stable ESU.

\section{The Einstein static solution and stability from the fix point approach}

By using equations (\ref{fr}) and (\ref{ma}) the Raychadhuri equation in the closed isotropic and homogeneous FLRW universe ($\kappa=+1$) can be written as\footnote{Here we have set units $8\pi G=1$.}
\begin{eqnarray}\label{aa}
[2(1-3\lambda)-6\omega\lambda]\ddot{a}&=&\frac{\dot{a}^{2}}{a}\left[-3(1+\omega)(1-4\lambda)+2(1-3\lambda)-6\omega\lambda\right]
\\\nonumber&&-\frac{1}{a}\left[(1-6\lambda)+3\omega(1-2\lambda)\right]+\frac{D}{a^{3}}(\omega-\omega_{\Lambda}).
\end{eqnarray}
The Einstein static solution is given by $\ddot{a} = 0 = \dot{a}$. To begin with we obtain the conditions
for the existence of this solution. The scale factor in this case is given by
\begin{eqnarray}
a^{2}_{_{Es}}=\frac{D(\omega-\omega_{\Lambda})}{(1-6\lambda)+3\omega(1-2\lambda)}.
\end{eqnarray}
The existence condition reduces to the reality condition for $a_{_{Es}}$, which for a positive $D$ takes
the forms
\begin{eqnarray}\label{ww}
\omega>\frac{-1+6\lambda}{3(1-2\lambda)} ~~~~{\rm and} ~~~\omega>\omega_{\Lambda},
\end{eqnarray}
or
\begin{eqnarray}\label{www}
\omega<\frac{-1+6\lambda}{3(1-2\lambda)} ~~~~{\rm and} ~~~\omega<\omega_{\Lambda}.
\end{eqnarray}

Here, we are going to study the stability of the critical point. For convenience,
we introduce two  variables
\begin{eqnarray}
x_1=a,\quad x_2=\dot{a}.
\end{eqnarray}
It is then easy to obtain the following equations
\begin{eqnarray}
\dot{x}_1=x_2,
\end{eqnarray}
\begin{eqnarray}
[2(1-3\lambda)-6\omega\lambda]\dot{x_2}&=&\frac{\dot{x_2}^{2}}{x_1}\left[-3(1+\omega)(1-4\lambda)+2(1-3\lambda)-6\omega\lambda\right]
\\\nonumber&&-\frac{1}{x_1}\left[(1-6\lambda)+3\omega(1-2\lambda)\right]+\frac{D}{x_{1}^{3}}(\omega-\omega_{\Lambda}).
\end{eqnarray}
According to these variables, the fixed point, $x_1=a_{Es},\; x_2=0$ describes the Einstein static solution properly. The stability of the critical
point is determined by the eigenvalue of the coefficient matrix ($J_{ij}=\frac{\partial \dot{x}_{i}}{\partial x_{j}} $)
stemming from linearizing the system explained in details by above two equations near the critical point. Using $\gamma^2$ to obtain the
eigenvalue we have
\begin{eqnarray}
\gamma^2=\frac{-A}{a_{Es}^{4}}(\omega-\omega_{\Lambda}).
\end{eqnarray}
In the case of $\gamma^{2}< 0$ the Einstein static solution has a center
equilibrium point, so it has circular stability, which means that small perturbation from the fixed
point results in oscillations about that point rather than exponential deviation from it. In
this case, the universe oscillates in the neighborhood of the Einstein static solution
indefinitely.
Thus, the stability condition is determined by $\gamma^{2}< 0$. For $A> 0$, this means
that $\omega>\omega_{\Lambda}$. Comparing this inequality with the conditions for existence of the Einstein
static solution, (\ref{ww}) and (\ref{www}), we find that the Einstein universe is stable for $\omega>\frac{-1+6\lambda}{3(1-2\lambda)}$.
Especially, it is stable in the presence of ordinary matter ($ \omega$) plus a vacuum term with negative pressure.
\begin{figure*}[ht]
    \includegraphics[width=2.5in]{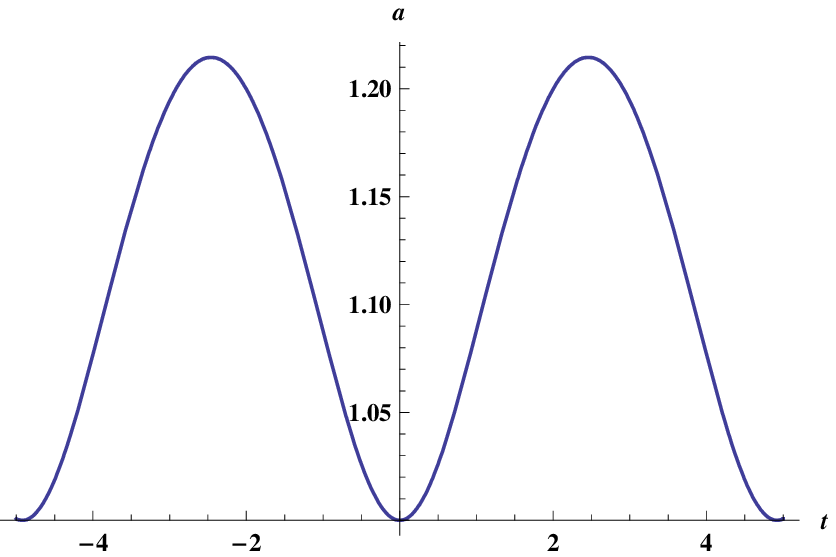}~~~~~~~~~~~~~~~
   \includegraphics[width=2.5in]{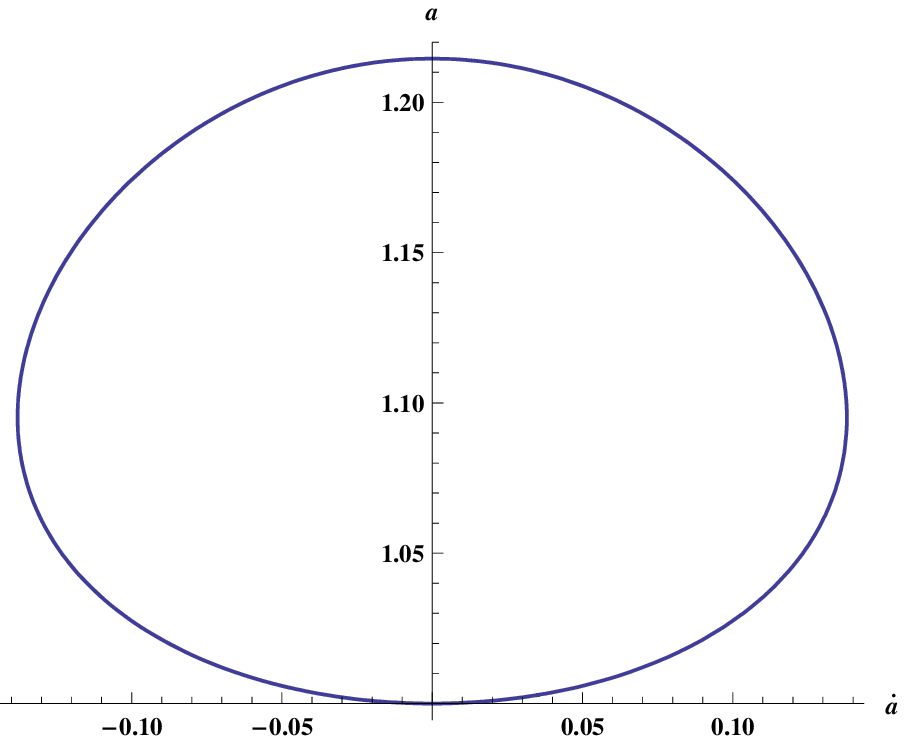}
 \caption{The evolutionary curve of the scale factor with time (left) and
the phase diagram  in space ($a$, $\dot{a}$) (right) for $\omega = 1/3$.}
  \label{stable}
\end{figure*}

To more study, we consider the effects of vacuum energy in the Rastall theory on the dynamics of the universe. As an example, we study the case where the energy content consists of vacuum energy, which we put $\omega_{\Lambda} = -1$ and $\rho_{\Lambda} = A/a^{4}$, in addition to
a relativistic matter with an equation-of-state parameter $\omega = 1/3$. Using these equation of
state parameters in equation (\ref{aa}) we obtain
\begin{eqnarray}
3a^{3}\ddot{a}+3a^{2}\dot{a}^{2}+3a^{2}-\frac{2A}{(1-4\lambda)}=0
\end{eqnarray}

From the above equation the corresponding scale factor of Einstein static solution is given by $a_{Es}^{2} = 2A/3(1-4\lambda)$.
Obviously, phase space trajectories which is beginning precisely on the Einstein static fixed point remain there indeterminately. From another point of view, trajectories which are creating in the vicinity
of this point would oscillate indefinitely near this solution.
An example of such a universe trajectory using initial conditions given by $a(0) = 1$ and $\dot{a}(0) = 0$,
with $A = 1.1$ and $\lambda=0.1$ has been plotted in Fig. 1.

\section{Conclusion}
We have discussed the existence and stability of the Einstein static universe in the presence of perfect fluid with a vacuum energy corresponding to conformally-invariant fields through
the Rastall theory of gravity. Using
the fix point method, we have found that the scale factor of Einstein static universe for closed deformed isotropic and homogeneous FLRW universe depends on the coupling parameter $\lambda$ between the energy-momentum tensor and the gradient of Ricci scalar. Also, we have determined the allowed intervals for the equation of state parameters related to the vacuum energy such that the Einstein universe is stable, while it is dynamically belonging to a center equilibrium point. The motivation for study of such a solution is  its essential role in
the construction of non-singular emergent oscillatory model which is past eternal, and hence can resolve
the singularity problem in the standard cosmological scenario.

\section*{Acknowledgments}
F. Darabi acknowledges the support of Azarbaijan Shahid Madani University  for the Sabbatical Leave, and thanks the hospitality of ICTP (Trieste)  during the Sabbatical Leave. K. Atazadeh acknowledges the financial support   by Research Institute for Astronomy and Astrophysics of
Maragha (RIAAM) under research project NO.1/4717-33.

\end{document}